\documentclass[conference]{IEEEtran}
\IEEEoverridecommandlockouts
\usepackage{balance}
\usepackage{amssymb}
\usepackage{stmaryrd}
\usepackage{color}
\usepackage{bbm}
\usepackage{multirow}
\usepackage{graphicx,times}
\usepackage{epstopdf}
\usepackage{indentfirst}
\usepackage{CJK}
\usepackage{amsmath}
\usepackage{amsfonts}
\usepackage{txfonts}
\usepackage{mathrsfs}
\usepackage{subfigure}
\usepackage{graphicx}
\usepackage{theorem}
\usepackage{url}
\usepackage{microtype}
\usepackage{fancyhdr}  

\def\BibTeX{{\rm B\kern-.05em{\sc i\kern-.025em b}\kern-.08em
		T\kern-.1667em\lower.7ex\hbox{E}\kern-.125emX}}

\begin{document}
	
\title{Towards Cross-Modal Forgery Detection and Localization on Live Surveillance Videos}
\author{\IEEEauthorblockN{Yong Huang{$^{^\dagger}$}, Xiang Li{$^{^\dagger}$}, Wei Wang{${^\ast}{^{^\dagger}}$}, Tao Jiang{$^{^\dagger}$}, Qian Zhang{${^\S}$}}\IEEEauthorblockA{{$^{^\dagger}$}School of Electronic Information and Communications, Huazhong University of Science and Technology\\ {${^\S}$} Department of Computer Science and Engineering, Hong Kong University of Science and Technology\\Email: \{yonghuang, xiangli\_ee, weiwangw, taojiang\}@hust.edu.cn, qianzh@cse.ust.hk}
	\thanks{${^\ast}$The corresponding author is Wei Wang (weiwangw@hust.edu.cn).}
}	

\maketitle

\pagestyle{fancy}
\thispagestyle{fancy}          
\fancyhead{}                     
\chead{IEEE INFOCOM 2021 - IEEE International Conference on Computer Communications}
\cfoot{} 
\renewcommand{\headrulewidth}{0pt}     
\renewcommand{\footrulewidth}{0pt}

\begin{abstract}
The cybersecurity breaches render surveillance systems vulnerable to video forgery attacks, under which authentic live video streams are tampered to conceal illegal human activities under surveillance cameras. Traditional video forensics approaches can detect and localize forgery traces in each video frame using computationally-expensive spatial-temporal analysis, while falling short in real-time verification of live video feeds. The recent work correlates time-series camera and wireless signals to recognize replayed surveillance videos using event-level timing information but it cannot realize fine-grained forgery detection and localization on each frame. To fill this gap, this paper proposes Secure-Pose, a novel cross-modal forgery detection and localization system for live surveillance videos using WiFi signals near the camera spot. We observe that coexisting camera and WiFi signals convey common human semantic information and the presence of forgery attacks on video frames will decouple such information correspondence. Secure-Pose extracts effective human pose features from synchronized multi-modal signals and detects and localizes forgery traces under both inter-frame and intra-frame attacks in each frame. We implement Secure-Pose using a commercial camera and two Intel 5300 NICs and evaluate it in real-world environments. Secure-Pose achieves a high detection accuracy of 95.1\% and can effectively localize tampered objects under different forgery attacks. 

\end{abstract}


\section{Introduction}
\begin{figure}
	\centering
	\includegraphics[width=0.9\linewidth]{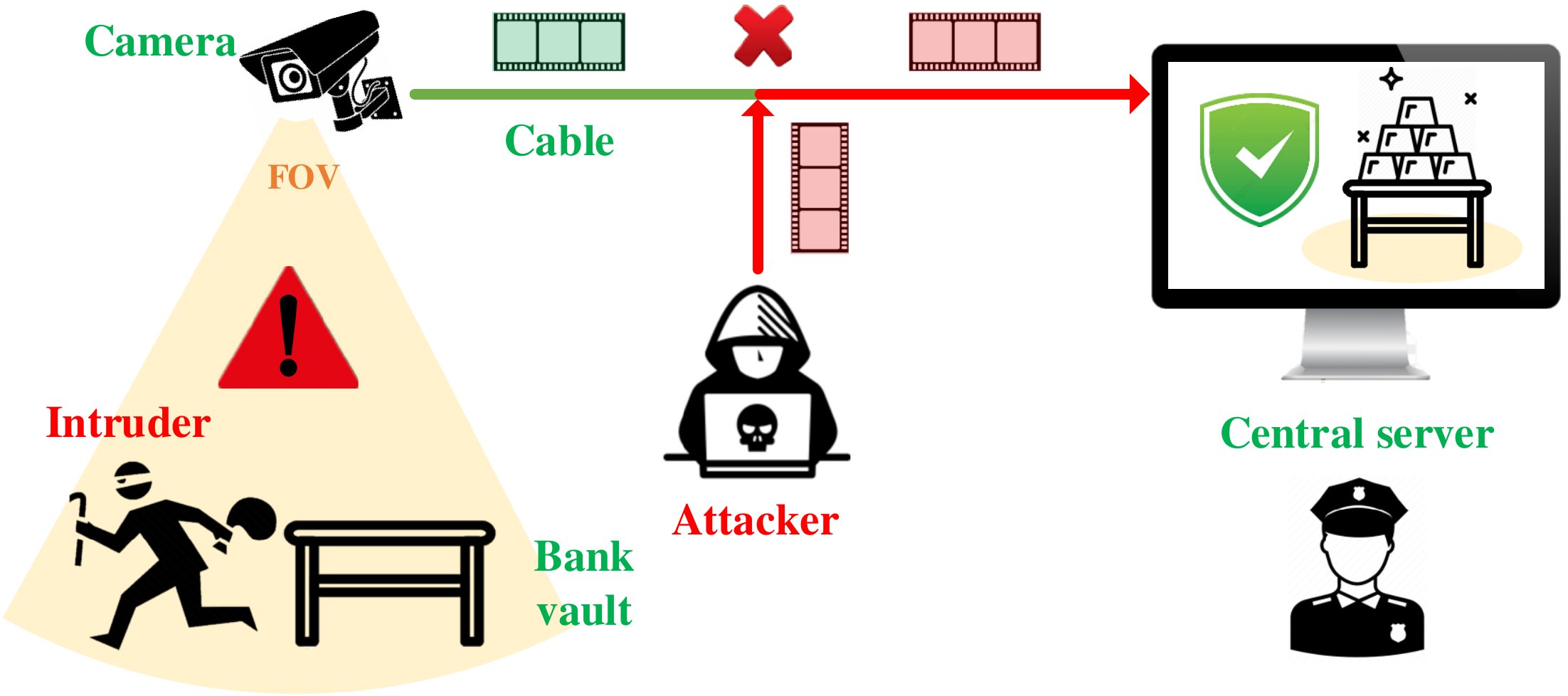}
	\caption{Illustration of video forgery attacks in surveillance systems.}
	\label{fig:attackscenario}
\end{figure}

With the increasing needs of safety and security in our daily life, video surveillance systems have gained a lot of traction in a wide spectrum of indoor applications, such as crime prevention in banks and customer monitoring in retail stores~\cite{liu2013intelligent}. As their popularity and prominence rapidly grow in the physical world, these systems inevitably become attractive attack surfaces in the cybersecurity space. Recent studies have demonstrated that attackers can infiltrate into the surveillance system by exploiting vulnerabilities of the monitoring camera~\cite{exploiting2013} or hijacking its connection Ethernet cable~\cite{looping2015} and then tamper the authentic live video streams to cover illegal human activities in the monitored area without showing any perceptible clues in the central server's screen as shown in Fig.~\ref{fig:attackscenario}. Under the looming threat of such attacks, timely forgery detection and accurate forgery localization on each video frame is highly desired for a surveillance system to quickly alarm on-going cyberattacks and seamlessly track potential intruders. 

Although extensive efforts have been devoted to detecting video forgery attacks, existing approaches still fall short in achieving both real-time and fine-grained detection performance. Traditional watermark-based approaches require dedicated modules on surveillance cameras for video integrity preservation, while not all camera manufacturers support such modules. Alternatively, many video forensics approaches that exploit statistic characteristics of video signals~\cite{fayyaz2020improved,yang2016using,chen2015automatic,ulutas2017frame,wang2007exposing} are developed to detect tampered frames and further localize forged regions in them. Though possessing fine-grained detection abilities, these approaches basically rely on various spatial-temporal analysis methods, which require a high computational complexity, and therefore are ill-suited for real-time attack detection on live surveillance videos. Additionally, the recent work~\cite{lakshmanan2019surfi} demonstrates that WiFi signals can be leveraged to expose video looping attacks on surveillance systems. However, it employs event-level timing information from time-series WiFi and camera signals to detect replayed video sequences and cannot realize fine-grained forgery detection and localization on each video frame. Hence, none of existing approaches simultaneously satisfies real-time and fine-grained requirements of forgery detection and localization.

The pervasive coexistence of surveillance cameras and WiFi devices offers the opportunity to detect video forgery attacks in a real-time and fine-grained manner. Nowadays, many areas under surveillance cameras, such as shops and homes, are also covered by WiFi hotspots to provide us ubiquitous wireless connectivity~\cite{wang2017sampleless,wang2020enabling}. In such areas, not only visible light but only WiFi signals interact with involving human objects, because human bodies act as reflectors in the WiFi frequency range. In this condition, camera and WiFi signals convey the common human semantic information. If forgery attacks are launched on video frames, such cross-modal information correspondence will be decoupled, which can be exploited for timely forgery detection and accurate forgery localization.

Toward this end, we present Secure-Pose, a novel cross-modal system that effectively detects and localizes video forgery traces in live surveillance video streams using coexisting WiFi signals. Our fundamental insight is that channel state information (CSI) measurements in WiFi signals are effective in probing human objects under surveillance cameras, from which authentic human semantic information can be carefully disentangled for forgery detection and localization on suspicious live video frames. Specifically, under video forgery attacks, the human semantic features from video frames are tampered to conceal really on-going scenes and thus will mismatch with that of CSI measurements. In this way, taking concurrent camera and WiFi signals as input, Secure-Pose can discern between authentic and tampered video frames and further localize forgery traces in each video frame.  

To realize the above idea, we have to address the following three challenges.

\textit{1) How to synchronize noisy CSI measurements with video frames?} Generally, raw CSI measurements have variable time intervals between them due to the random access protocol and packet loss, which significantly hampers periodicity of CSI measurements and their accordance with video frames. To deal with this issue, we first use linear interpolation to resample a group of fixed-interval CSI samples between two video frames based on CSI and video timestamps. Next, we leverage Hampel identifier to remove unwanted outliers from CSI samples caused by environment noise.

\textit{2) How to effectively disentangle human semantic information from CSI measurements?} Due to a low spatial resolution, it is highly challenging to disengage semantic information from CSI measurements. To address this issue, we first formulate semantic feature extraction as pose estimation and present pose information using Joint Heat Maps (JHMs) and Part Affinity Fields (PAFs) that are  retrievable from CSI samples. We then leverage two customized neural networks that take camera and WiFi as inputs, respectively, with an effective cross-modal training scheme to output JHMs and PAFs.

\textit{3) How to efficiently detect and localize forgery traces in each video frame?} Since multiple people would appear in the camera's FOV, it is cumbersome and computationally inefficient to perform person-by-person attack detection based on estimated JHMs and PAFs from camera and WiFi signals. To avoid this issue, we construct the JHM difference tensor that can preserve sufficient information about forgery attacks while removing irrelevant information. Based on such feature tensor, we can detect video forgery attacks using a sample convolutional neural network (CNN) and localize human objects only in tampered regions.

\textbf{Summary of Results.} We design Secure-Pose that extracts effective human pose features from synchronized camera and CSI signals and then performs forgery detection and localization on each video frame. We implement Secure-Pose using a logitech 720p camera and two Intel 5300 NICs, and evaluate it in real-world indoor environments. The evaluation results demonstrate that Secure-Pose achieves a high detection accuracy of 95.1\%. Moreover, it can successfully recognize 94.9\% of tampered video frames and meanwhile mistakenly classify just 4.7\% authentic ones. In addition, Secure-Pose can localize body keypoints of tampered human objects with a mPCK@0.2 of 69.3\% under various forgery attacks.

\textbf{Contributions.} The main contributions of this work are summarized as follows. First, we show that coexisting camera and WiFi signals convey common human semantic information and such cross-modal information correspondence can be exploited to detect and localize video forgery attacks in each video frame. Second, we propose a novel cross-modal forgery detection and localization system that can discover forgery traces in live surveillance videos. Third, we implement Secure-Pose on a commercial logitech 720p camera and two Intel 5300 NICs and conduct extensive experiments in real-world indoor environments to verify its effectiveness and robustness against various forgery attacks.

\section{Threat Model and WiFi CSI Signatures}\label{sec:motivation}

\begin{figure}
	\centering
	\subfigure[Inter-frame attacks.]{
		\includegraphics[width=0.475\linewidth]{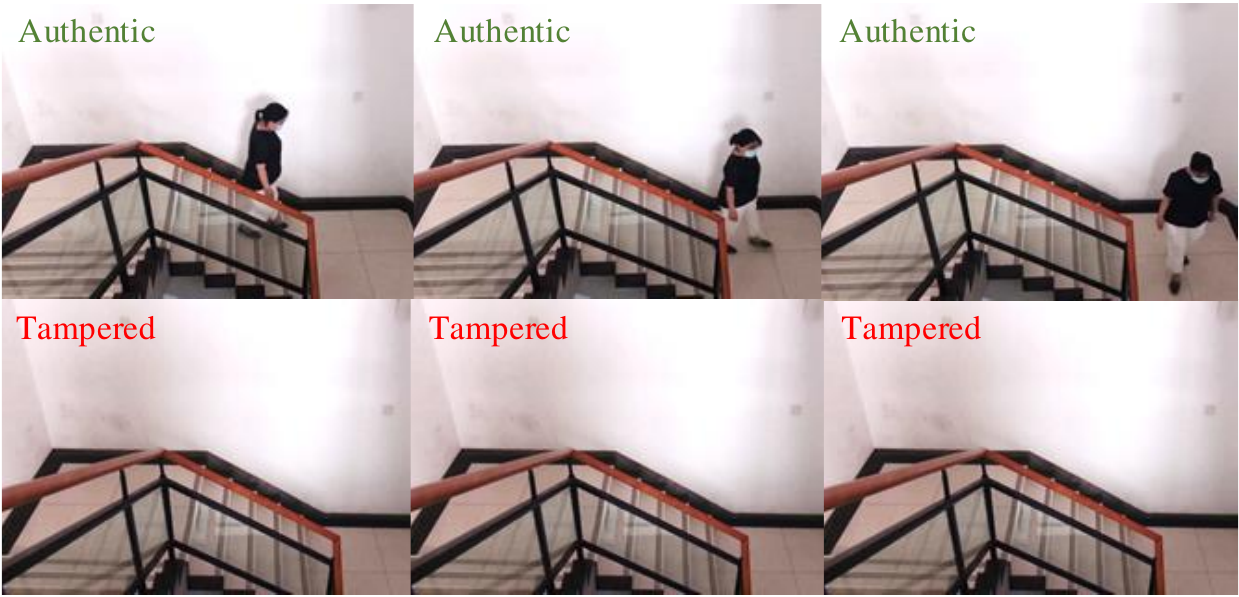}}
	\label{1a}
	\subfigure[Intra-frame attacks.]{
		\includegraphics[width=0.475\linewidth]{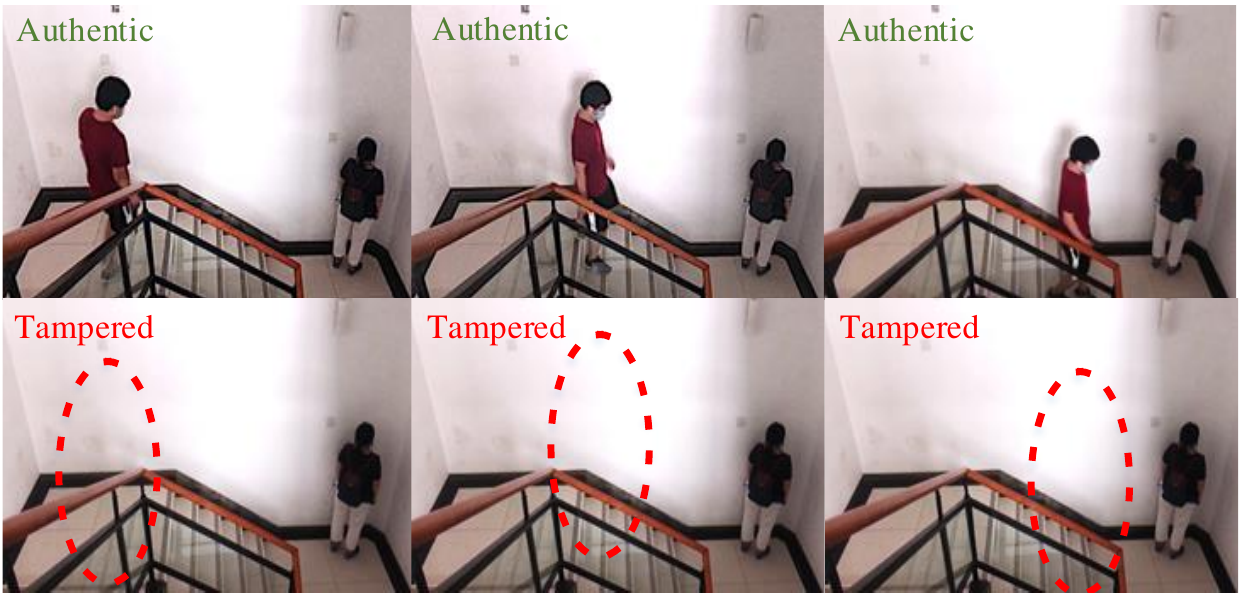}}
	\label{1b}\\
	\caption{Examples of inter-frame and intra-frame attacks.}
	\label{fig:attackexample} 
\end{figure}

\subsection{Video Forgery Attacks in Surveillance Systems}
We consider a common surveillance scenario, where a fixed camera monitors an open area in an indoor environment, such as banks and retail stores, and transmits live video streams to a central server for remote monitoring. In this scenario, human objects are generally the target of surveillance systems for behavior monitoring and trace tracking~\cite{liu2013intelligent}. Moreover, since many people with different motions could go in and out of the monitored area in reality, we make no assumption on the number of people in the camera's field-of-view (FOV) as well as their body motions. 

In such a surveillance system, a malicious attacker could be capable of launching various cyberattacks, such as hijacking the camera~\cite{exploiting2013} or the connection Ethernet cable~\cite{looping2015}, to penetrate into the system. After that, the attacker can further trigger two kinds of video forgery attacks, i.e., inter-frame attacks and intra-frame attacks, to conceal illegal human activities under surveillance cameras. In inter-frame attacks, the attacker replaces the live video frames with previously-recorded frames. In intra-frame attacks, the attacker removes or adds some human objects in the transmitted frames. Fig.~\ref{fig:attackexample} exemplifies inter-frame and intra-frame attacks on sequences of surveillance video frames. In general, both two kinds of video forgery attacks render the human semantic features, such as body positions and poses, in the tampered video frames and really on-going scenes mismatching. The video frames that describe mismatching human activities are considered as tampered frames, and the others are authentic frames. 

Furthermore, we assume that a pair of WiFi transceivers are colocated with the surveillance camera. Since it is extremely difficult to mimic complicated CSI measurements based on video contents, CSI measurements from the WiFi transceivers are considered to be authentic.

\subsection{CSI Signatures in Commercial WiFi Signals}
In WiFi communication systems, channel state information describes how a WiFi signal propagates between a transceiver pair for characterizing current channel conditions. Specifically, according to the IEEE 802.11n WiFi protocol~\cite{xiao2005ieee}, both orthogonal frequency division multiplexing (OFDM) and multiple-input-multiple-output (MIMO) technologies are adopted for high-throughput transmission. Let $ N_{tx} $ and $ N_{rx} $ be the numbers of transmitting and receiving antennas, respectively, $ K $ the number of OFDM subcarriers. At time $ t $, one CSI sample between the $ i $-th transmitting antenna and the $ j $-th receiving antenna on $ k $-th subcarrier can be expressed as~\cite{wang2015understanding}
\begin{align} \label{eq: csi}
	h_{i,j,k}^{t}=\sum^{N_L}_{n=1} a_n e^{j2\pi \frac{d^t_{i,j,n}}{\lambda_k}},
\end{align}
where $ N_L $ represents the number of signal traveling paths, $ d^t_{i,j,n} $ the signal travel distance on $ n $-th path and $ \lambda_k $ the wavelength of the $ k $-th subcarrier. Thus, with the adoption of OFDM and MIMO technologies, a CSI measurement can be denoted as $ \mathbf{H}^t \in \mathbb{C}^{ N_{tx}\times N_{rx} \times K}$. From Eq.~\eqref{eq: csi}, we can observe that one CSI sample consists of signals from multiple paths. 

Since the human body can be considered as a reflector at WiFi frequency range, WiFi signals will have rich interactions, such as reflection and scattering, with body limbs before arriving at the receiver if human objects are present near the WiFi transceivers. Hence, similar to visible light captured by a camera, WiFi CSI measurements can also convey human body information, which can be leveraged to verify the veracity of video content under the threat of video forgery attacks.

\section{System Design}\label{sec:design}

\begin{figure}
	\centering
	\includegraphics[width=0.9\linewidth]{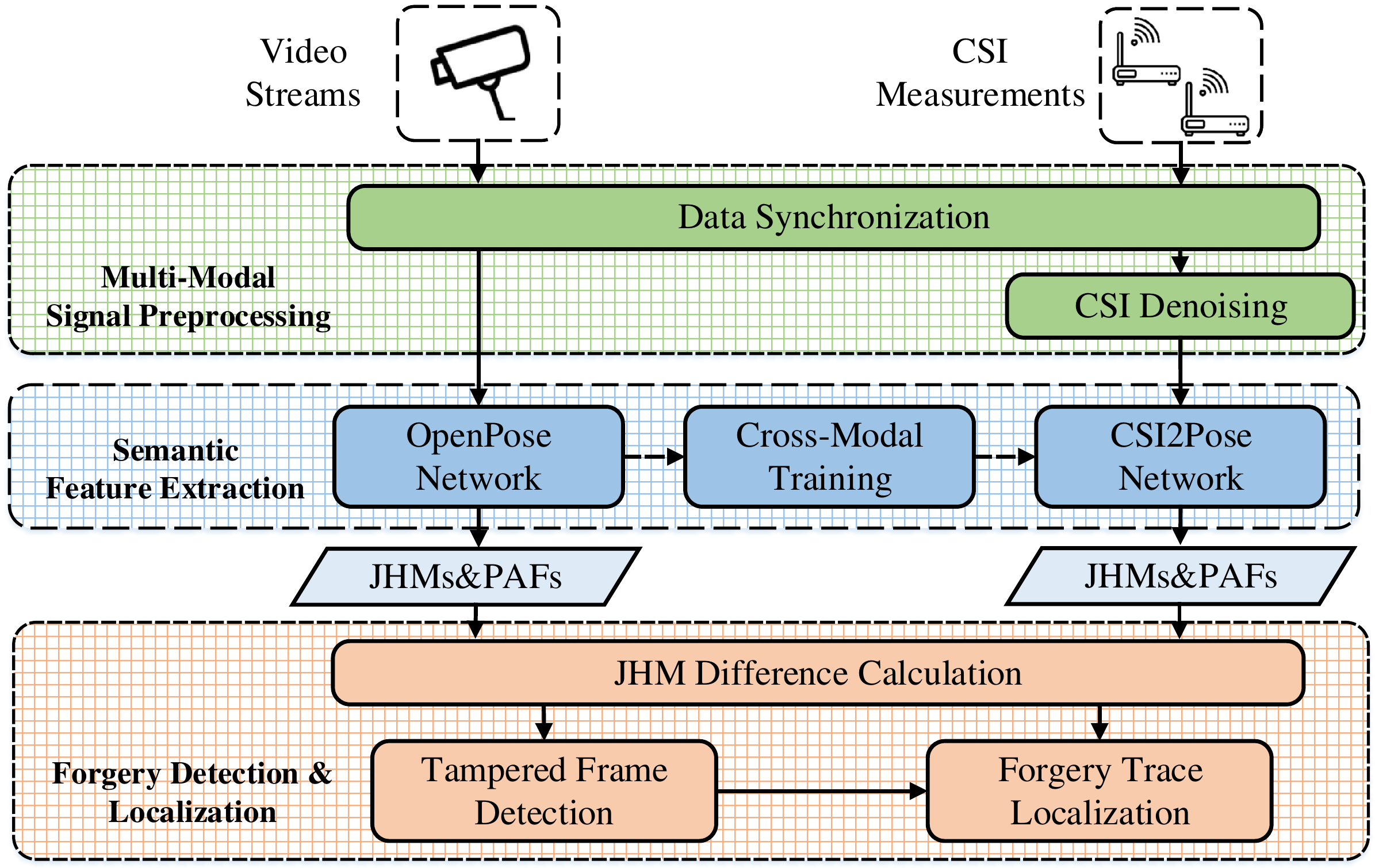}
	\caption{System architecture of Secure-Pose.}
	\label{fig:systemflow}
\end{figure}

\subsection{System Overview}

Secure-Pose takes advantage of pervasive WiFi signals in the monitored area to detect and localize forgery traces in live surveillance videos. In particular, it enables real-time forgery detection as well as fine-grained forgery localization on each video frame. Secure-Pose can be a part of the surveillance system, and its properties empower the system to quickly alarm on-going cyberattacks and seamlessly track potential intruders. 

As depicted in Fig.~\ref{fig:systemflow}, Secure-Pose takes concurrent camera and WiFi signals as input and outputs the authenticity of live video frames and potential forgery traces. The core of our system includes three components -- \textit{Multi-Modal Signal Preprocessing}, \textit{Semantic Feature Extraction} and \textit{Forgery Detection and Localization}.

\begin{itemize}
	\item  \textbf{Multi-Modal Signal Preprocessing.} Taking both camera and WiFi signals as input, our system first synchronizes CSI amplitude features with video frames via linear interpolation for better data accordance. Then, it removes outliers from noisy amplitudes using Hampel identifier.
	\item  \textbf{Semantic Feature Extraction.} In this component, our system exploits well-designed OpenPose~\cite{cao2017realtime} and CSI2Pose neural networks, respectively, to extract both Joint Heat Maps and Part Affinity Fields from synchronous video frame and CSI features. Moreover, a cross-modal training scheme is developed to effectively train CSI2Pose network with the supervision of OpenPose network. 
	\item  \textbf{Forgery Detection and Localization.} In this component, our system first calculates the difference of JHMs from OpenPose and CSI2Pose networks for efficient feature representation. Then, it builds a simple detection network to accurately detect forgery attacks. After that, our system localizes forgery traces in each tampered frame.  
\end{itemize}

\subsection{Multi-Modal Signal Preprocessing}

Being a part of the surveillance system, Secure-Pose first takes as input compressed video streams transmitted from the surveillance camera and decode them as RGB image frames. Specifically, we denote the decoded video frame sequence as
\begin{align}
\left\lbrace \cdots, \mathbf{I}^{m-1}, \mathbf{I}^{m}, \mathbf{I}^{ m+1},\cdots \right\rbrace,
\end{align}
where $ \mathbf{I}^{m} \in \mathbb{R}^{H\times W \times 3} $ is a complete RGB image. Therein, $ H $ and $ W $ indicate the frame height and width, respectively. In parallel with video data acquisition, we also input raw CSI amplitude measurements from the co-located WiFi receiver into our system. Formally, the input amplitude sequence can be expressed as
\begin{align}
\left\lbrace \cdots, \mathbf{A}^{n-1}, \mathbf{A}^{n}, \mathbf{A}^{n+1},\cdots \right\rbrace,
\end{align}
where $ \mathbf{A}^{n} \in \mathbb{R}^{N_{tx} \times N_{rx} \times K}  $ contains all amplitudes of $  \mathbf{H}^{n} $.

\textbf{Data Synchronization.} Since Secure-Pose relies on concurrent camera and WiFi signals, the synchronization between two signals is critical. The asynchronous signals probably convey different human semantic information, leading to a high false alarming rate when be applied for forgery detection.

\begin{figure}
	\centering
	\subfigure[Time intervals between successive CSI measurements.]{
		\includegraphics[width=0.43\linewidth]{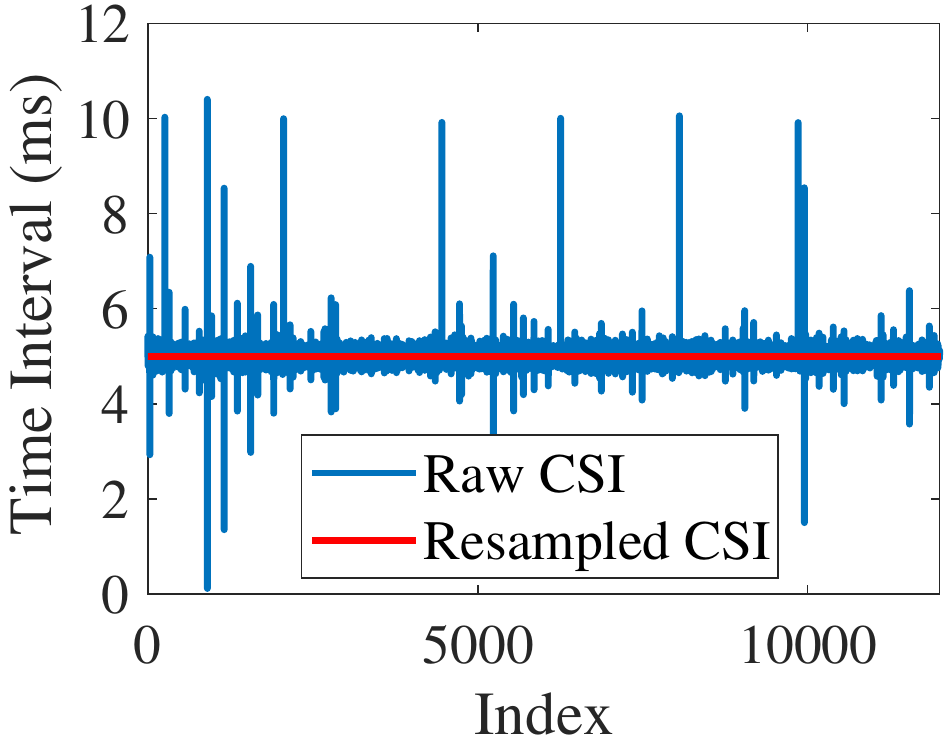}}
	\label{2a}
	\subfigure[CSI measurements of one subcarrier with detected outliers.]{
		\includegraphics[width=0.43\linewidth]{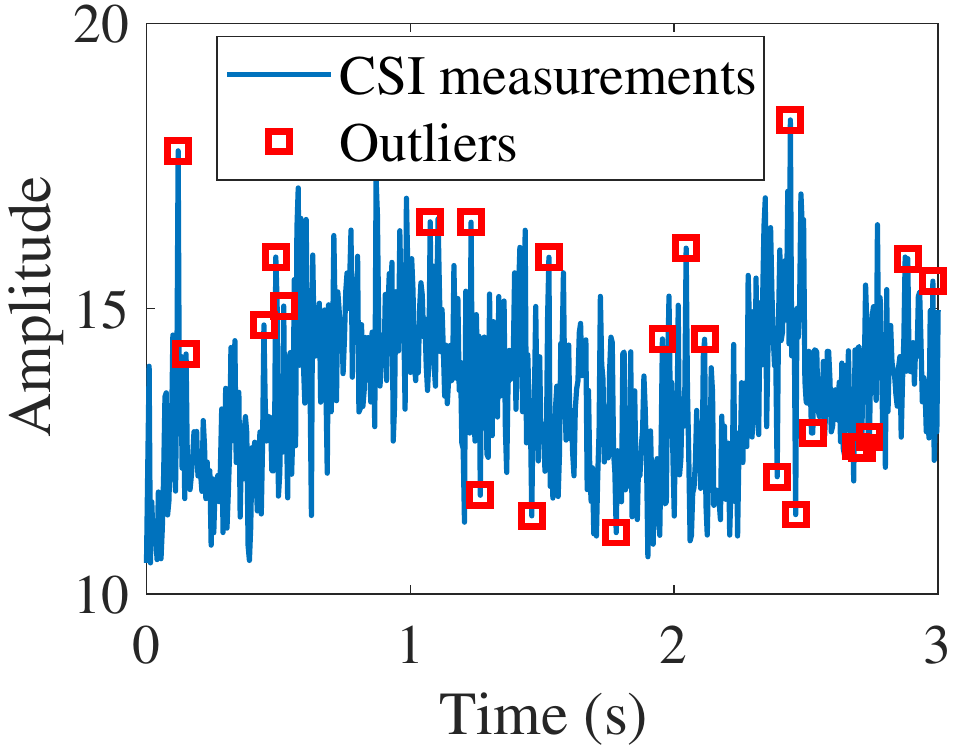}}
	\label{2b}\\
	\caption{CSI measurements synchronization and denoising.}
	\label{fig:datapreprocessing} 
\end{figure}

Concretely, let us assume that the camera has a FPS (Frames Per Second) of $ F_{I} $ and the WiFi receiver has a CSI sampling rate of $ F_{W} $ Hz. Because $ F_{W} $ is basically much larger than $ F_{I} $, we correspond $ F \le   F_{W}/F_{I}  $ CSI measurements to one video frame. However, due to the random access protocol as well as packet loss, the time interval between any two successive raw CSI measurements is variable as shown in Fig.~\ref{fig:datapreprocessing}(a), which results in a varying number of CSI measurements in one time unit and thus weakens their periodicity. In contrary, the time interval between video frames is basically constant. Consequently, simply correlating the CSI measurement index $ n $ with the video frame index $ m $ is erroneous. To deal with this issue, our system synchronizes CSI measurements with video frames based on their timestamps. Mathematically, we denote $ t_{m-1} $ and $ t_m $ as the timestamps of two video frames $ \mathbf{I}^{m-1}$ and $ \mathbf{I}^{m} $, respectively. For the video frame $ \mathbf{I}^{m} $, our system resamples a set of $ F $ CSI measurements $ \left\lbrace \mathbf{\bar{A}}^{1}, \cdots, \mathbf{\bar{A}}^{f}, \cdots, \mathbf{\bar{A}}^{F}  \right\rbrace  $ with a constant time interval $ \Delta t = (t_{m}-t_{m-1})/F $ using low-complexity linear interpolation, and thus each resampled CSI measurement $ \mathbf{\bar{A}}^{f} $ can be obtained by
\begin{align} 
\mathbf{\bar{A}}^{f} = \mathbf{A}^{n-1} + \eta \cdot (\mathbf{A}^{n}-\mathbf{A}^{n-1}),
\end{align}
where $ t_{n-1} \leq t_f = t_{m-1} + f \Delta t \leq t_{n} $ and $ \eta = \frac{t_{f} - t_{n-1}}{t_{n} - t_{n-1}}$. As Fig.~\ref{fig:datapreprocessing}(a) shows, the resampled CSI measurements have a constant periodicity and thus better accordance with video frames. 

\textbf{CSI Denoising.} Due to environment noise, outliers that show a sudden change could appear in CSI measurements and thus hamper the effectiveness of extracted amplitude features. To address this problem, we leverage Hampel identifier~\cite{davies1993the}, a simple yet efficient outlier detection algorithm, to remove these outliers over time domain. In particular, given a CSI amplitude sequence $ \left\lbrace a_{i,j,k}^{n-\delta} \cdots , a_{i,j,k}^{n},\cdots,a_{i,j,k}^{n-\delta} \right\rbrace  $ of $ i $-th transmitting antenna, $ j $-th receiving antenna and $ k $-th subcarrier, our system declares $ a_{i,j,k}^{n} $ as an outlier such that
 \begin{align}\label{eq:hampel filter}
 	\left| a_{i,j,k}^{n} - \mu^n \right| >  \gamma \cdot \sigma^n,
 \end{align}
where $ \mu^n $ and $ \sigma^n $, respectively, are the local median and median absolute deviation of the sequence. Moreover, $ \delta $ determines the sliding window size and $ \gamma $ decides the tolerable deviation from $ \mu^n $, and we empirically set both $ \delta $ and $  \gamma $ to be 3 in our system. Once $ a_{i,j,k}^{n} $ is detected as an outlier, our system replaces it with the local median $ \mu^n $. As shown in Fig.~\ref{fig:datapreprocessing}(b), most of outliers in CSI measurements can be effectively detected. 
  
Towards this end, at time $ t_m $, Secure-Pose outputs a set of $ F $ resampled and denoised CSI amplitudes, which can be denoted as $ \mathbf{R}^{m} $. Similar to the video frame $ \mathbf{I}^{m} $, we term $ \mathbf{R}^{m} $ as the RF frame. Thus, our system obtains a camera-WiFi frame pair $ (\mathbf{I}^m,\mathbf{R}^m) $ after multi-modal signal preprocessing.

\subsection{Semantic Feature Extraction} 

With a camera-WiFi frame pair $ (\mathbf{I}^m,\mathbf{R}^m) $, Secure-Pose proceeds to extract semantic features from them, respectively.

\textbf{Semantic Feature Representation.} We formulate the task that extracts semantic features as a human pose estimation problem. In computer vision, human pose estimation requires correctly localizing anatomical keypoints or body parts of individuals and further inferring their poses in images. Such limb-level information is perceptible by WiFi signals as human limbs act as reflectors at WiFi frequency ranges. 

In the field of pose estimation, there are mainly top-bottom and bottom-top approaches, which correspond to different ways of representing human pose information. Specifically, the top-bottom approaches~\cite{he2017mask,ren2015faster} first employ a person detector to segment all persons out with bounding boxes on pixel-level feature maps and then perform person-wise pose estimation within their own bounding boxes. In these approaches, the estimated human poses are represented by locations of body keypoints within the corresponding bounding boxes. However, these bounding-box-based approaches cannot be directly applied to low-spatial-resolution RF frames, from which pixel-level visual feature maps are extremely hard to extract. In contrast, the bottom-top approaches~\cite{pishchulin2016DeepCut,cao2017realtime} first extract all possible body keypoints in one image and associate them to form full-body poses. In this way, human poses are represented by locations of body keypoints in the global image space. Thus, the bottom-top pose representation does not rely on pixel-level bounding boxes, which is more suitable for extracting human pose features from RF frames. 

Concretely, we represent human pose features as Joint Heat Maps (JHMs) and Part Affinity Fields (PAFs)~\cite{cao2017realtime}. We assume there are $ J $ body keypoints. Given the video frame $  \mathbf{I}^{m} \in \mathbb{R}^{H\times W \times 3} $, JHMs indicate the confidence maps of body keypoint locations in the image space and can be represented by a 3D tensor $ \mathbf{S}^{m} \in \mathbb{R}^{H\times W \times J} $ as
\begin{align}
	\mathbf{S}^{m} = \left( \mathbf{s}_1^m, \mathbf{s}_2^m, \cdots, \mathbf{s}_J^m \right), 
\end{align}
where $ \mathbf{s}_j^m \in \mathbb{R}^{H\times W} $ is the 2D confidence map of $ j $-th keypoint. In our system, we use the Body-14 model and thus set $ J=14 $. Moreover, PAFs contain spatial information of body limbs, each of which is represented by two associated body keypoints. Formally, PAFs of $ C $ limbs can be denoted as a 4D tensor $ \mathbf{L}^m  \in \mathbb{R}^{H\times W \times 2 \times C} $ as
\begin{align}
	\mathbf{L}^m = \left( \mathbf{l}_1^m, \mathbf{l}_2^m, \cdots, \mathbf{l}_C^m \right),
\end{align} 
where $ \mathbf{l}_c^m  \in \mathbb{R}^{ H\times W \times 2 }$ is a set of 2D vectors at all elements in the image space and indicates the location and orientation for $ c $-th limb. Since the Body-14 model is adopted, we have $ C=13 $.

\begin{figure}
	\centering
	\includegraphics[width=0.9\linewidth]{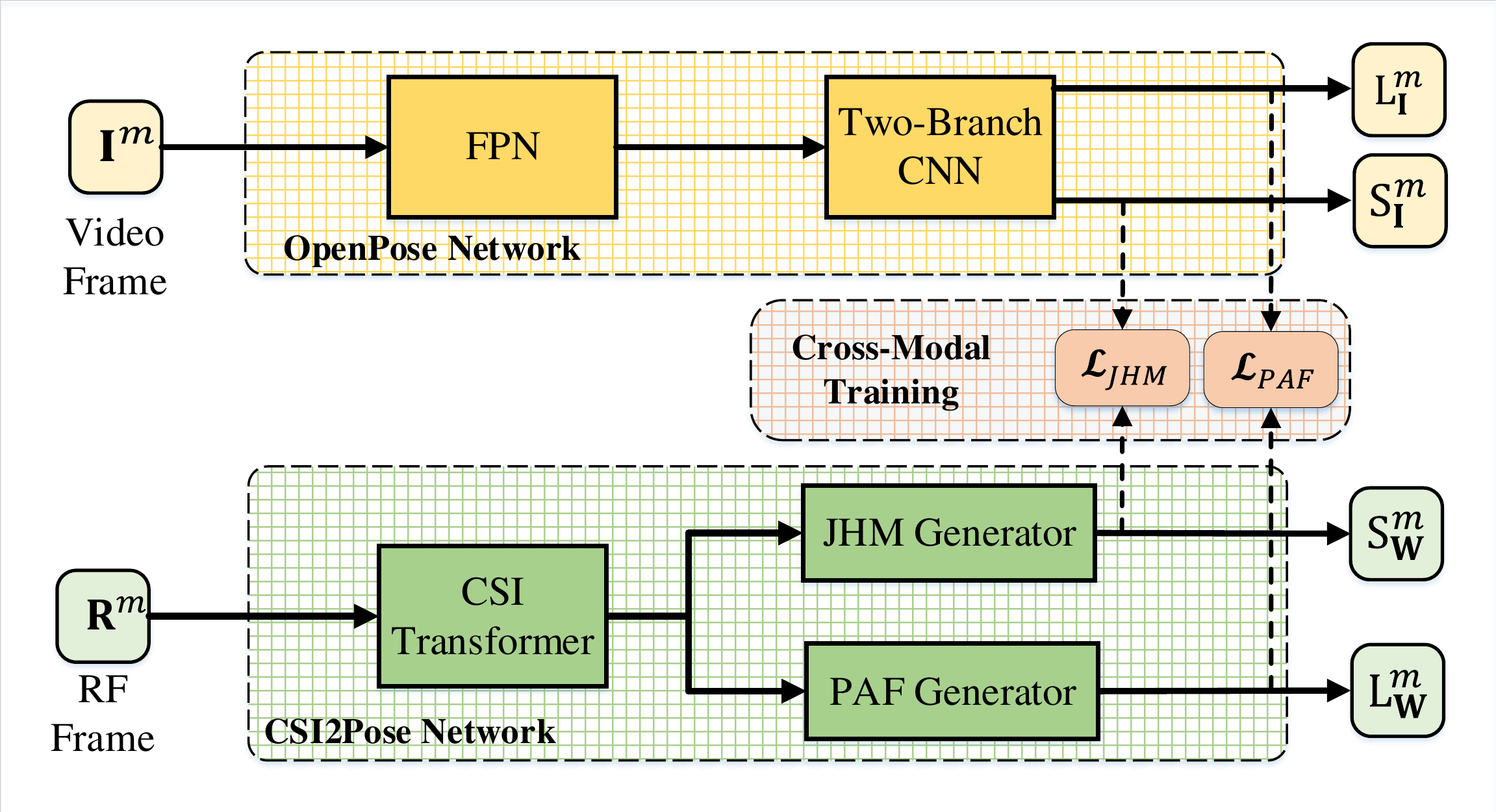}
	\caption{The work flow of semantic feature extraction. }
	\label{fig:network-architecture}
\end{figure}

As Fig.~\ref{fig:network-architecture} shows, we exploit \textit{OpenPose} network~\cite{cao2017realtime} to extract JHMs and PAFs from  $  \mathbf{I}^m $ and develop \textit{CSI2Pose} network for $  \mathbf{R}^m $.

\textbf{OpenPose Network.} Taking $ \mathbf{I}^m $ as input, it first adopts a Feature Pyramid Network (FPN)~\cite{lin2017feature}, a state-of-the-art feature extraction tool to learn multi-scale visual features from images. Then, it leverages a two-branch CNN to generate visual JHMs and PAFs. Thus, given the input video frame $ \mathbf{I}^m $,it outputs JHMs $ \mathbf{S}^m_{\mathbf{I}} \in \mathbb{R}^{H \times W \times J} $ and PAFs $ \mathbf{L}^m_{\mathbf{I}} \in \mathbb{R}^{H \times W \times 2 \times C } $ as
\begin{align}
	\left( \mathbf{S}^m_{\mathbf{I}}, \mathbf{L}^m_{\mathbf{I}} \right)  =  \mathcal{F_I}(\mathbf{I}^m),
\end{align}
where $ \mathcal{F_I}(\cdot) $ represents the parameters of OpenPose network.

\textbf{CSI2Pose Network.}  Given the RF frame $ \mathbf{R}^m $, our CSI2Pose network generates corresponding JHMs $ \mathbf{S}^m_{\mathbf{W}} \in \mathbb{R} ^{H \times W \times J} $ and PAFs $ \mathbf{L}^m_{\mathbf{W}} \in \mathbb{R}^{H \times W \times C \times 2} $. Since it is a challenging task to translate time-series CSI amplitudes to human pose features, we design three modules in CSI2Pose network, i.e., CSI Transformer, JHM Generator and PAF Generator as depicted in Fig~\ref{fig:network-architecture}.

In CSI transformer, we transform $ \mathbf{R}^m  $ into an image-size tensor that encodes human semantic information under the camera' FOV. To do this, we first use deconvolution layers to upsample $ \mathbf{R}^m $ to a higher spatial resolution. Then, we employ two convolution layers to produce encoded wireless features and feed them into a stack of residual blocks~\cite{he2016deep} to further transform the wireless features into intermediate feature maps that correlate with human information in the image space.

In the subsequent JHM generator, the intermediate feature maps are fed into a decoder structure to refine semantic features. Next, a Fully Convolutional Network (FCN)~\cite{long2015fully} is followed to map refined semantic features into the JHM tensor $ \mathbf{S}^m_{\mathbf{W}} $. Similarly, we build the same structure for our PAF generator to output the PAF tensor $ \mathbf{L}^m_{\mathbf{W}} $.

After all, given the RF frame $ \mathbf{R}^m  $, our CSI2Pose network outputs JHMs $ \mathbf{S}^m_{\mathbf{W}} $ and PAFs $ \mathbf{L}^m_{\mathbf{W}} $ as
\begin{align}
	\left( \mathbf{S}^m_{\mathbf{W}}, \mathbf{L}^m_{\mathbf{W}} \right)  =  \mathcal{F_W}(\mathbf{R}^m),
\end{align}
where $ \mathcal{F_W}(\cdot) $ are trainable parameters of CSI2Pose network.

\textbf{Cross-Modal Training.} In the training phase, we train CSI2Pose network with the supervision of OpenPose network's outputs for avoiding laborious and time-consuming data annotation. Specifically, given a training set of camera-WiFi frame pairs $ \left\lbrace (\mathbf{I}^y,\mathbf{R}^y)  \right\rbrace_{y=1:Y}  $, we first input $ \left\lbrace \mathbf{I}^y \right\rbrace_{y=1:Y}  $ into OpenPose network $ \mathcal{F_{I}} (\cdot) $ and obtain the corresponding visual semantic feature set $ \left\lbrace  (\mathbf{S}^y_{\mathbf{I}}, \mathbf{L}^y_{\mathbf{I}}) \right\rbrace_{y=1:Y}  $. After that, we proceed to train our CSI2Pose network by taking $ \left\lbrace \mathbf{R}^y \right\rbrace_{y=1:Y} $ as input and using $ \left\lbrace  (\mathbf{S}^y_{\mathbf{I}}, \mathbf{L}^y_{\mathbf{I}}) \right\rbrace_{y=1:Y}  $ as ground-truth labels. 

The training objective of CSI2Pose network is to minimize the discrepancy between its output and OpenPose's output:
\begin{align}\label{eq: training object}
\mathop{\min} \limits_{\mathcal{F_{W}}} \sum_{y=1}^{Y} \mathcal{L}_{JHM} \left( \mathbf{S}^y_{\mathbf{I}}, \mathbf{S}^y_{\mathbf{W}} \right) + \mathcal{L}_{PAF} \left( \mathbf{L}^y_{\mathbf{I}}, \mathbf{L}^y_{\mathbf{W}} \right) ,
\end{align}
where $ \mathcal{L}_{JHM} (\cdot,\cdot) $ and $ \mathcal{L}_{PAF} (\cdot,\cdot) $ are mean squared error (MSE) loss functions for JHM and PAF features, and they can be further expressed as
\begin{align}
	\mathcal{L}_{JHM} \left( \mathbf{S}_{\mathbf{I}}, \mathbf{S}_{\mathbf{W}} \right)= \sum_{j=1}^{J} \sum_{h,w} \alpha^j_{h,w} \cdot || \mathbf{s}^j_{\mathbf{I}} (h,w) - \mathbf{s}^j_{\mathbf{W}} (h,w)||^2_2, \label{eq:loss jhm} \\
	\mathcal{L}_{PAF} \left( \mathbf{L}_{\mathbf{I}}, \mathbf{L}_{\mathbf{W}} \right)= \sum_{c=1}^{C} \sum_{h,w} \alpha^c_{h,w} \cdot || \mathbf{l}^c_{\mathbf{I}} (h,w) - \mathbf{l}^c_{\mathbf{W}} (h,w) ||^2_2. \label{eq:loss paf}
\end{align}
In Eq.~\eqref{eq:loss jhm} and Eq.~\eqref{eq:loss paf}, $ \alpha^j_{h,w} $ and $ \alpha^c_{h,w} $ are pixel-wise weights for JHM and PAF tensors, respectively. Since the majority of elements in JHMs and PAFs are with small values~\cite{cao2017realtime}, we set both $ \alpha^j_{h,w} $ and $ \alpha^c_{h,w} $ to be proportional to the absolute value of $ (h,w) $-th element in the ground-truth labels $ \mathbf{S}_{\mathbf{I}} $ and $ \mathbf{S}_{\mathbf{W}} $ for giving positive elements more attentions. In addition, considering that JHMs and PAFs indicate different semantic information and have values in different scales, we appoint different coefficients $ \lambda_1 $,  $ \beta_1 $ , $ \lambda_2 $ and $ \beta_2 $ to balance between $ \mathcal{L}_{JHM} $ and $ \mathcal{L}_{PAF} $ in the training objective~\eqref{eq: training object}. Hence, $ \alpha^j_{h,w} $ and $ \alpha^c_{h,w} $ can be computed as
\begin{align}
	\alpha^j_{h,w} = \lambda_1 \cdot ||\mathbf{s}^j_{\mathbf{I}}(h,w)||^2_2 + \beta_1, \label{eq:weighted jhm loss}\\
	\alpha^c_{h,w} = \lambda_2 \cdot ||\mathbf{l}^c_{\mathbf{I}}(h,w)||^2_2 + \beta_2. \label{eq:weighted paf loss}
\end{align}
In our system, we set $\lambda_1=1$, $\beta_1=1$,  $\lambda_2=0.3$ and $\beta_2=0.7$.

\subsection{Forgery Detection and Localization.}

\begin{figure}
	\centering
	\includegraphics[width=0.95\linewidth]{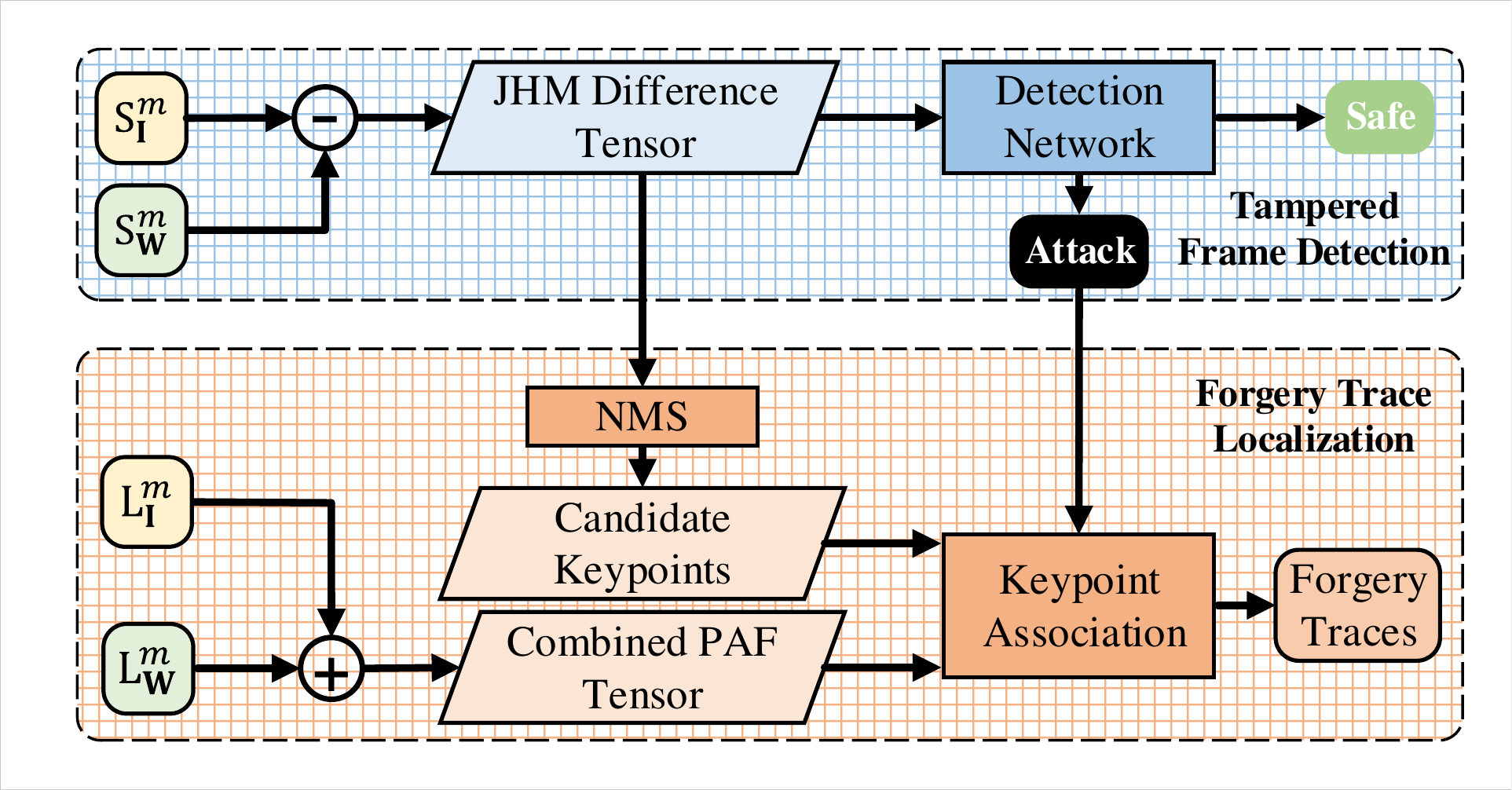}
	\caption{The work flow of video forgery detection and localization.}
	\label{fig:detectionlocalization}
\end{figure}

We proceed to detect forgery attacks and localize forgery traces in $ \mathbf{I}^m $ based on $ \mathbf{S}^m_{\mathbf{I}} $, $ \mathbf{L}^m_{\mathbf{I}} $, $ \mathbf{S}^m_{\mathbf{W}} $ and $ \mathbf{L}^m_{\mathbf{W}} $ as shown in Fig.~\ref{fig:detectionlocalization}.

\textbf{JHM Difference Calculation.} Since we have JHMs and PAFs estimated from video and RF frames, a straightforward way is to assemble detected keypoints to form human instances with full-body poses from $ \left( \mathbf{S}^m_{\mathbf{I}}, \mathbf{L}^m_{\mathbf{I}} \right) $ and $ \left( \mathbf{S}^m_{\mathbf{W}}, \mathbf{L}^m_{\mathbf{W}} \right) $, respectively, and perform forgery detection via calculating pose similarity in a person-wise manner. However, such approach is cumbersome and computationally inefficient.

Secure-Pose utilizes estimated JHMs $ \mathbf{S}^m_{\mathbf{I}} $ and $ \mathbf{S}^m_{\mathbf{W}} $ for forgery detection. Specifically, it takes their difference as the basis of the following forgery detection and localization. Let us denote $ \mathbf{D}^m \in \mathbb{R}^{H \times W \times J} $ as the JHM difference tensor, which can be computed as
\begin{align}\label{eq:jhm residual}
	\mathbf{D}^m = \mathbf{S}^m_{\mathbf{I}} - \mathbf{S}^m_{\mathbf{W}}. 
\end{align}
The rationale behind the difference operation in Eq.~\eqref{eq:jhm residual} is that with effective feature extraction by $ \mathcal{F_I}(\cdot) $ and $ \mathcal{F_W}(\cdot) $, JHM tensors $ \mathbf{S}^m_{\mathbf{I}} $ and $ \mathbf{S}^m_{\mathbf{W}} $ from video and RF frames would be very similar, making $ \mathbf{D}^m $ having a lot of near-zero differences, if no attacks are present. In contrast, when video forgery attacks appear, body poses in the video frame $ \mathbf{I}^m $ is modified and the visual JHMs $ \mathbf{S}^m_{\mathbf{I}} $ will change accordingly, which results in large differences between $ \mathbf{S}^m_{\mathbf{I}} $ and $ \mathbf{S}^m_{\mathbf{W}} $ in the tampered regions. Hence, the JHM difference tensor $ \mathbf{D}^m $ can preserve sufficient information about forgery attacks and remove redundant information about authentic scenes. 

\begin{figure}
	\centering
	\includegraphics[width=0.9\linewidth]{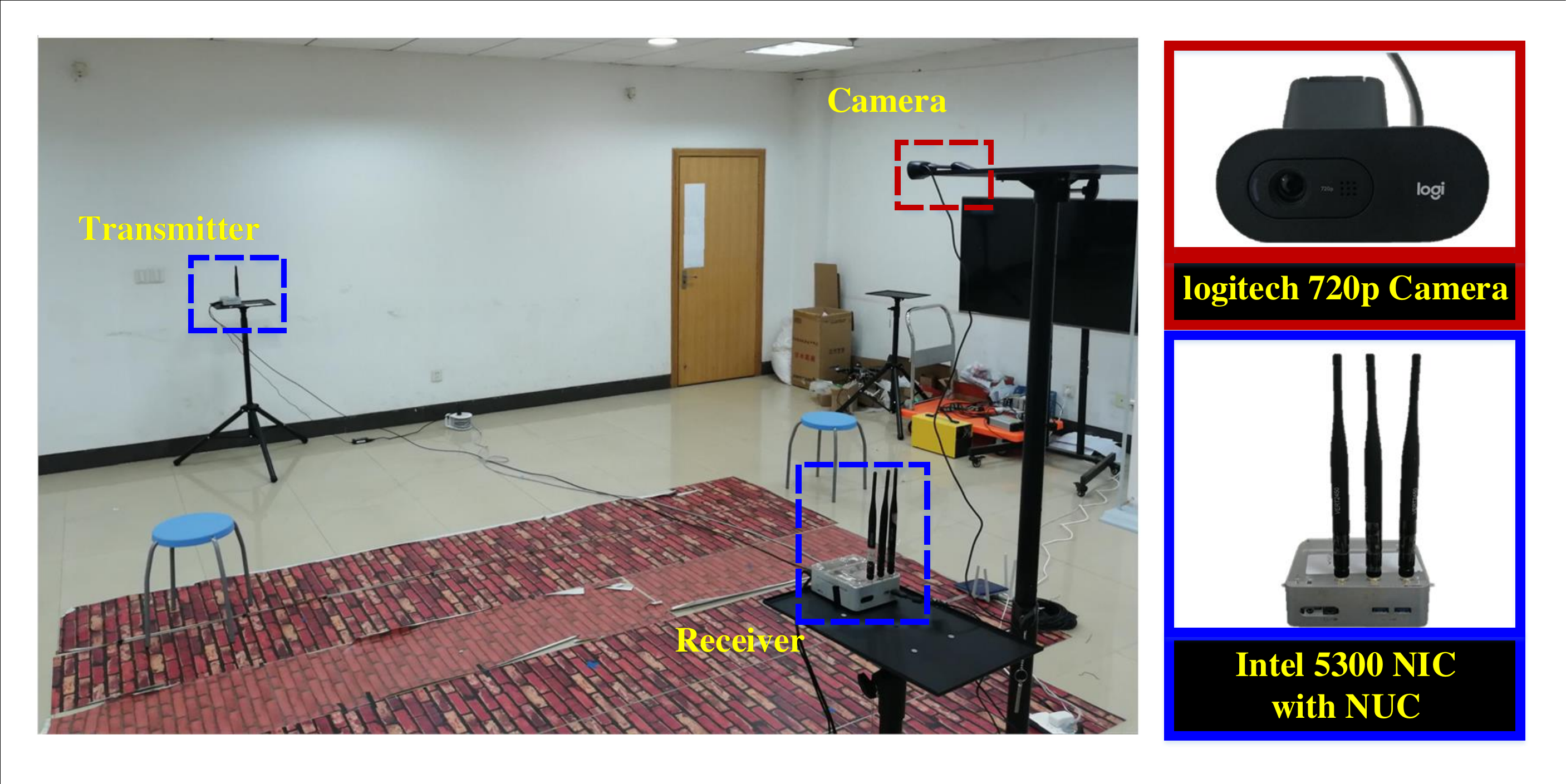}
	\caption{Experimental platform. It contains one logitech 720p camera and two Intel 5300 NICs.}
	\label{fig:experimentalsetup}
\end{figure}

\textbf{Tampered Frame Detection.} To avoid cumbersome person-by-person detection, we perform frame forgery detection using a simple binary classification model, which directly takes the JHM difference tensor $ \mathbf{D}^m $ as input and produces, as output, a binary indicator $ \mathcal{Z}^m \in \left\lbrace 0,1  \right\rbrace $ to mark the presence of forgery attacks in the video frame $ \mathbf{I}^m $. Note that we assign $ \mathcal{Z}^m = 1 $ to a positive $ \mathbf{D}^m $ if the corresponding $ \mathbf{I}^m $ is tampered and $ \mathcal{Z}^m = 0 $ to a negative $ \mathbf{D}^m $ if $ \mathbf{I}^m $ is authentic. 

Specifically, we build a detection neural network $ \mathcal{F_D} (\cdot) $ to recognize forgery attacks based on $ \mathbf{D}^m $. Specifically, $ \mathcal{F_D} (\cdot) $ has a CNN structure, which consists of two convolutional (conv) layers and two fully connected (FC) layers. The conv part first executes convolutions on all channels of $ \mathbf{D}^m $ and further converts features into a 1D feature vector as output. Then, the FC part transforms the 1D feature vector into a two-dimensional probability vector $ \mathbf{p}^m =   (p^m_0,p^m_1) \in \left[ 0,1 \right] ^{2} $ to indicate the likelihood of attack presence as
\begin{align}
	\mathbf{p}^m = \mathcal{F_D} (\mathbf{D}^m).
\end{align}
Once $ \mathbf{p}^m $ is obtained, our system can output a decision $ \mathcal{Z}^m $ as
\begin{align}
	\mathcal{Z}^m = \arg \mathop{\max} \limits_{z} p^m_z.
\end{align}

To train the detection network $ \mathcal{F_D} (\cdot) $, we collect lots of positive and negative samples of JHM difference tensors and thus obtain a labeled dataset $ \left\lbrace (\mathbf{D}^y, \mathcal{Z}^y) \right\rbrace_{y=1:Y} $. Hence, the training objective is to minimize the difference between the detection network's predictions and ground-truth labels:
\begin{align}\label{eq: training objective}
\mathop{\min} \limits_{\mathcal{F_{D}}} \frac{1}{Y} \sum_{y=1}^{Y} \mathcal{L}_{BCE} \left(\mathbf{p}^y, \mathcal{Z}^y  \right),
\end{align}
where $ \mathcal{L}_{BCE} (\cdot,\cdot) $ is binary cross entropy (BCE) loss, which can be computed as
\begin{align}
	\mathcal{L}_{BCE} \left(\mathbf{p}, \mathcal{Z}  \right) = -\left( \mathcal{Z} \cdot \log p_1 + \left( 1-\mathcal{Z}\right) \cdot \log p_0 \right). 
\end{align}

\textbf{Forgery Trace Localization.} After frame forgery detection, our system proceeds to localize forgery traces in the tampered frames. Since the tampered frames could contain more human objects or less human objects when compared with the authentic ones, we term both the added and erased human objects in tampered frames as abnormal objects for simplicity. Hence, the goal of forgery trace localization is to estimate human poses of abnormal objects in tampered regions. To do this, once $ \mathbf{I}^m $ is detected as a tampered frame by $ \mathcal{F_D} (\cdot) $, Secure-Pose performs body part association to discover poses of abnormal objects based on the JHM difference tensor $ \mathbf{D}^m $ as well as estimated PAF tensors $ \mathbf{L}^m_{\mathbf{I}} $ and $ \mathbf{L}^m_{\mathbf{W}} $. 

The first step is to select body keypoint candidates of abnormal objects based on $ \mathbf{D}^m $. Since $ \mathbf{D}^m $ is the difference between $ \mathbf{S}^m_{\mathbf{I}} $ and $ \mathbf{S}^m_{\mathbf{W}} $, it could have negative values as well as positive ones. Our system performs non-maximum suppression (NMS) on the absolute value of $ \mathbf{D}^m $ to pick up a keypoint candidate set $ \mathcal{K}^m $, which can be expressed as
\begin{align}
	\mathcal{K}^m =\left\lbrace \mathbf{k}^m_{j,n} \in \mathbb{R}^2 : \text{for} j \in \left\lbrace 1,\cdots ,J  \right\rbrace, n \in \left\lbrace 1, \cdots, N_j \right\rbrace  \right\rbrace,
\end{align}
where $ N_j $ denotes the number of candidates of $ j $-th keypoint and $ \mathbf{k}^m_{j,n}  $ is the location of $ n $-th candidate of $ j $-th keypoint. 

The second step is to associate the candidate keypoints $ \mathcal{K}^m $ to form abnormal human poses using PAFs  $ \mathbf{L}^m_{\mathbf{I}} $ and $ \mathbf{L}^m_{\mathbf{W}} $. In this step, our system first sums up  $ \mathbf{L}^m_{\mathbf{I}} $ and $ \mathbf{L}^m_{\mathbf{W}} $ to obtain a combined PAF tensor $ \mathbf{F}^m_{\mathbf{L}} \in \mathbb{R}^{H\times W \times 2 \times C} $ for saving computational complexity in the following association phase, which can be computed as
\begin{align}\label{eq: combined paf}
	\mathbf{F}^m_{\mathbf{L}} = \mathbf{L}^m_{\mathbf{I}} + \mathbf{L}^m_{\mathbf{W}}.
\end{align}
Then, with $ \mathcal{K}^m $ and $ \mathbf{F}^m_{\mathbf{L}} $, our system leverages the keypoint association method proposed in~\cite{cao2017realtime} to estimate the poses of abnormal objects. Specifically, human poses are determined by a set of connected body keypoits, and such best connection states $ \mathcal{E}^m  $ between keypoints in $ \mathcal{K}^m $ can be obtained by 
\begin{align}
	\mathcal{E}^m = \mathcal{F_A} \left( \mathcal{K}^m, \mathbf{F}^m_{\mathbf{L}} \right), 
\end{align}
where $  \mathcal{F_A} (\cdot,\cdot) $ represents the association function in~\cite{cao2017realtime}. Moreover, each element in $ \mathcal{E}^m  $ is a binary variable $ E_{k_1,k_2} \in \left\lbrace 0,1 \right\rbrace  $, which indicates the connection state between $ k_1 $-th keypoint and $ k_2 $-th keypoint in $ \mathcal{K}^m $. 

Toward this end, Secure-Pose outputs a binary detection decision $ \mathcal{Z}^m $ using a simple neural network $ \mathcal{F_{D}} (\cdot) $ and produces the body poses $ \mathcal{E}^m  $ of only abnormal objects in the tampered region.

\section{Implementation and Evaluation}\label{sec:implementation}

\subsection{Implementation}

As shown in Fig.~\ref{fig:experimentalsetup}, we implement Secure-Pose using one logitech 720p camera and two Intel 5300 NICs equipped with three antennas each. Specifically, two NICs, one as a transmitter and the left as a receiver, are both placed at a height of one meter and separated with each other about six meters away. In each NIC, we uniformly spaced three antennas with a distance of 2.6cm. For a good FOV, we mount the camera with a height of 2m and place it on the receiver side. To generate video signals, the camera is set to output 1280$\times$720p RGB images with a FPS of 7.5. To sample CSI signals, two NICs are controlled by two Intel NUCs separately, using an open-source tool~\cite{Halperin2011csitool}, to communicate with a bandwidth of 20 MHz centering in the 5.6 GHz WiFi band. In this condition, we record CSI measurements of 30 subcarriers with a sampling rate of 100Hz.

\subsection{Evaluation Methodology}

\textbf{Data Collection.} We collect concurrent camera and WiFi data in a laboratory office with a size of 8m$ \times $16m. We recruit five volunteers and ask them to perform daily activities, such as walking, sitting and waving hands, under the camera' FOV. During data collection, the number of people in the scene varies from zero to three. In such condition, we collect multi-modal data of half an hour in total. Based on the collected data, we resize all video frames into a size of $64\times 128$, and resample 5 CSI measurements for each video frame. After that, we obtain about 12K samples of synchronized video-RF fame pairs. In all samples, half of them are used to generate positive samples that contain tampered video frames under inter-frame or intra-frame attacks. To do this, we perform inter-frame attacks by randomly replacing the pristine video frame in a sample with a different video frame from another sample. To launch intra-frame attacks, we leverage Faster-RCNN~\cite{ren2015faster} to detect and crop a human object out and then replace it with the corresponding blank background segment. The left 6K samples with authentic video frames are considered as negative samples. Finally, we randomly split all positive and negative samples into three datasets for training and testing our system.
\begin{itemize}
	\item \textbf{Dataset A.} It contains 3K positive and 3K negative samples for training $ \mathcal{F_{W}} $ and $ \mathcal{F_{D}} $ in our system.
	\item \textbf{Dataset B.} It contains 1.5K positive samples under inter-frame attacks and 1.5K negative samples for testing.
	\item \textbf{Dataset C.} It contains 1.5K positive samples under intra-frame attacks and 1.5K negative samples for testing.
\end{itemize}

\textbf{Training Details.} We train our CSI2Pose and detection network on PyTorch framework. Specifically, we adopt a batch size of 1 and a RMSprop optimizer with weight decay 1e-8 and momentum 0.9 for each training. The initial learning rate is set to 1e-6 for CSI2Pose network and 1e-5 for the detection network. Moreover, the learning rate will be multiplied by a decay factor 0.1 if validation loss does not decrease for 5 epochs. Finally, we train CSI2Pose network with 20 epochs and the detection network with 10 epochs.

\textbf{Evaluation Metrics.} We use the following metrics to evaluate the performance of Secure-Pose.
\begin{itemize}
	\item \textbf{Accuracy.} It is defined as the ratio of the number of samples that are correctly detected to the total number of positive and negative samples.
	\item \textbf{True positive rate (TPR).} It is the ratio of the number of positive samples that are successfully detected to the total number of positive samples. 
	\item \textbf{False positive rate (FPR).} It is the ratio of the number of negative samples that are mistakenly recognized to the total number of negative samples.
	\item \textbf{Percentage of correct keypoint (PCK).} It is the ratio that the normalized distance from prediction $ \mathbf{x}^{j}_{p} $ to ground-truth $ \mathbf{y}^{j}_p $ of $ j $-th keypoint of $ p $-th person is less than $ \rho  $:
	\begin{align}
	\notag	\text{PCK}_j@\rho = \frac{1}{P} \sum_{p=1}^{P} \mathbbm{1} \left\lbrace \frac{||\mathbf{x}^{j}_{p}-\mathbf{y}^{j}_p||_2}{b^p} \le \rho \right\rbrace  ,
	\end{align}
	where $ 0< \rho < 1 $, $ P $ is the number of people and $ b^p $ is the diagonal length of $ p $-th person's bounding box. Moreover, the mean PCK over all keypoints is denoted as $ \text{mPCK}@\rho  $.
	\item \textbf{Dice loss.} Given two numerical tensors $ \mathbf{P} \in \mathbb{R}^3 $ and $ \mathbf{Q} \in \mathbb{R}^3 $, it is computed as 
	\begin{align}
	\notag	D(\mathbf{P}, \mathbf{Q})=1- \frac{2  \sum_{x,y,z} | p(x,y,z)| \cdot |q(x,y,z)|  }{ \sum_{x,y,z} | p(x,y,z)| +  |q(x,y,z)|}.
	\end{align}
\end{itemize}

\begin{figure}[t]
	\hfill
	\begin{minipage}[t]{0.47\linewidth}
		\centering
		\includegraphics[width=0.95\linewidth]{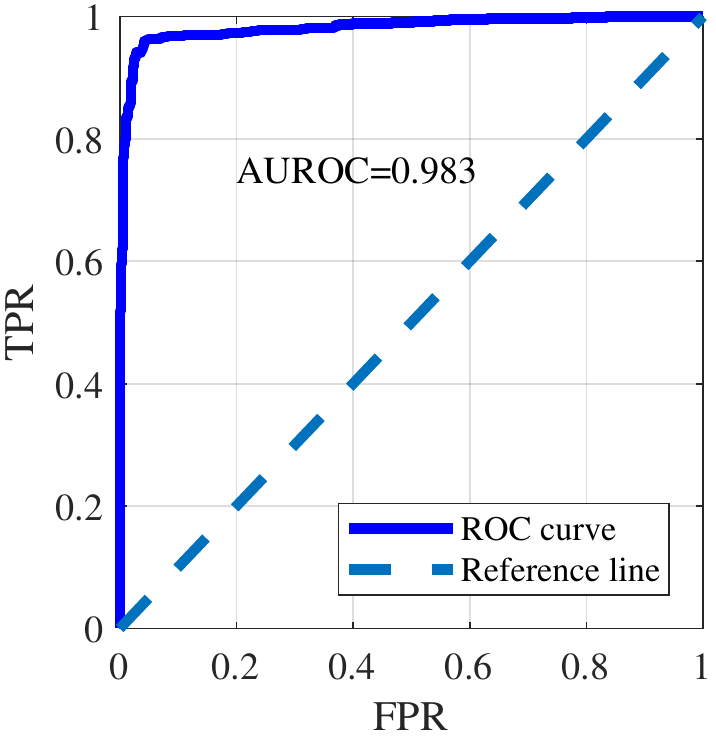}
		\caption{ROC curve. The reference line is for random guessing.}
		\label{fig:roccurve}
	\end{minipage}
	\hfill
	\begin{minipage}[t]{0.47\linewidth}
		\centering
		\includegraphics[width=0.97\linewidth]{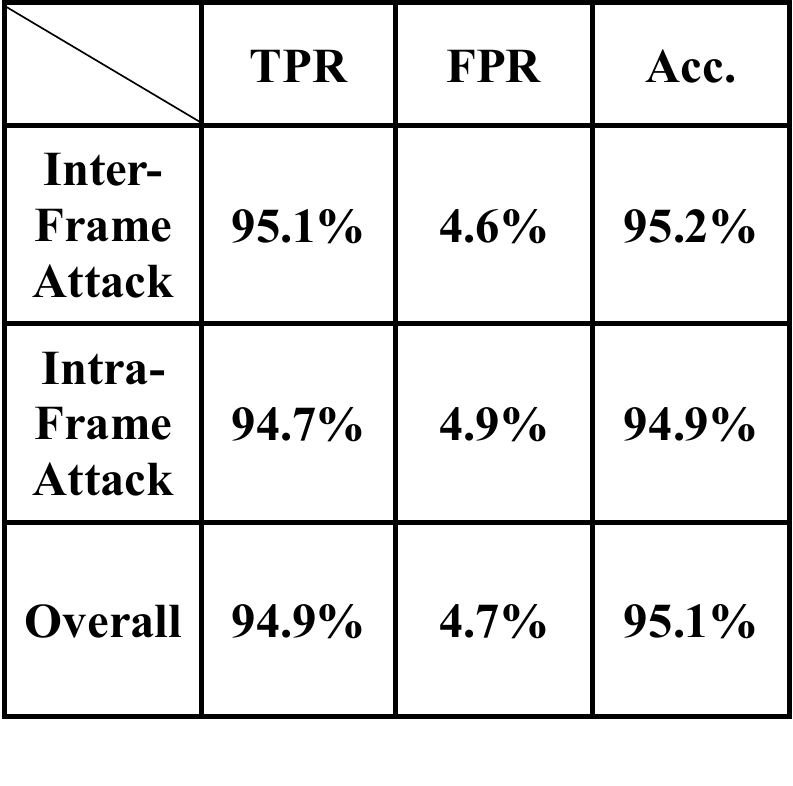}
		\caption{Forgery detection performance under different attacks.}
		\label{fig:overallperformance}
	\end{minipage}
\end{figure}

\begin{figure*}
	\hfill
	\begin{minipage}[t]{0.295\linewidth}
		\centering
		\includegraphics[width=0.98\linewidth]{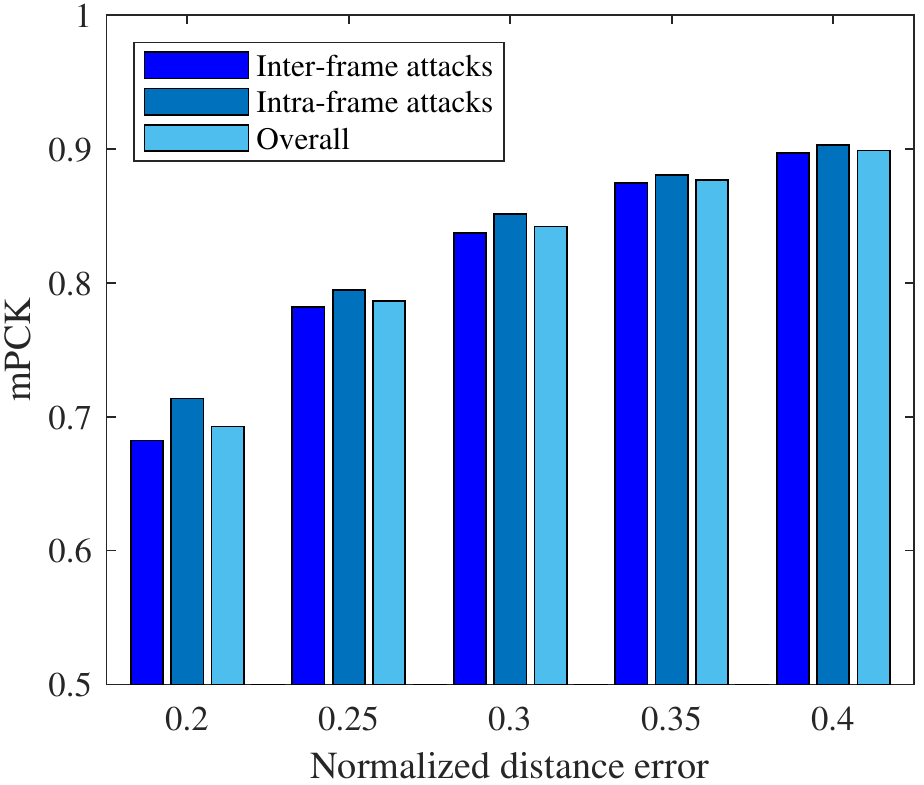}
		\caption{Pose estimation performance of abnormal objects in tampered regions.}
		\label{fig:forgerylocalizationperformance}
	\end{minipage}
	\hfill
	\begin{minipage}[t]{0.385\linewidth}
		\centering
		\includegraphics[width=1\linewidth]{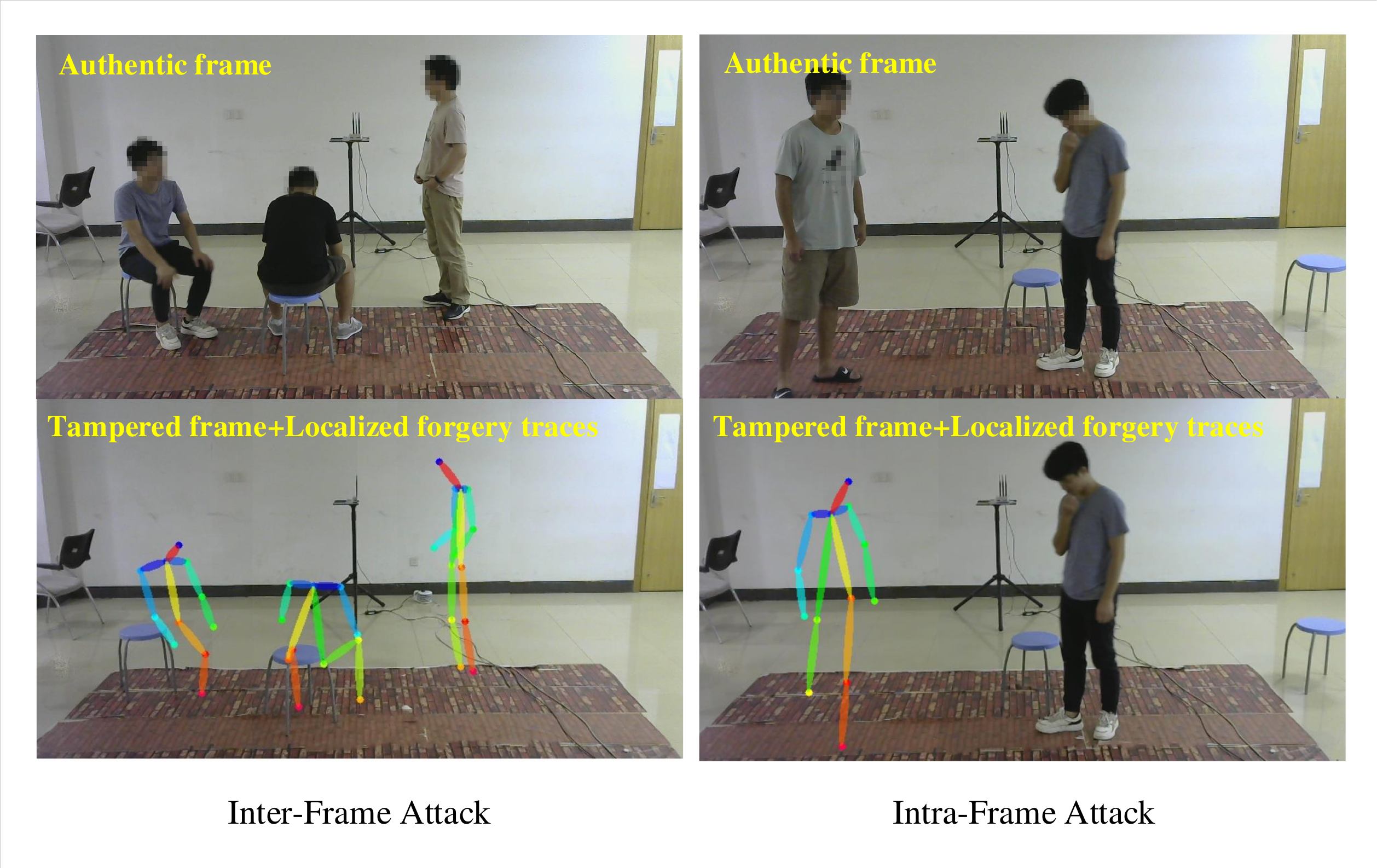}
		\caption{Instance of forgery trace localization under different attacks. All abnormal objects are successfully localized.}
		\label{fig:instanceinterframe}
	\end{minipage}
	\begin{minipage}[t]{0.295\linewidth}
		\centering
		\includegraphics[width=0.99\linewidth]{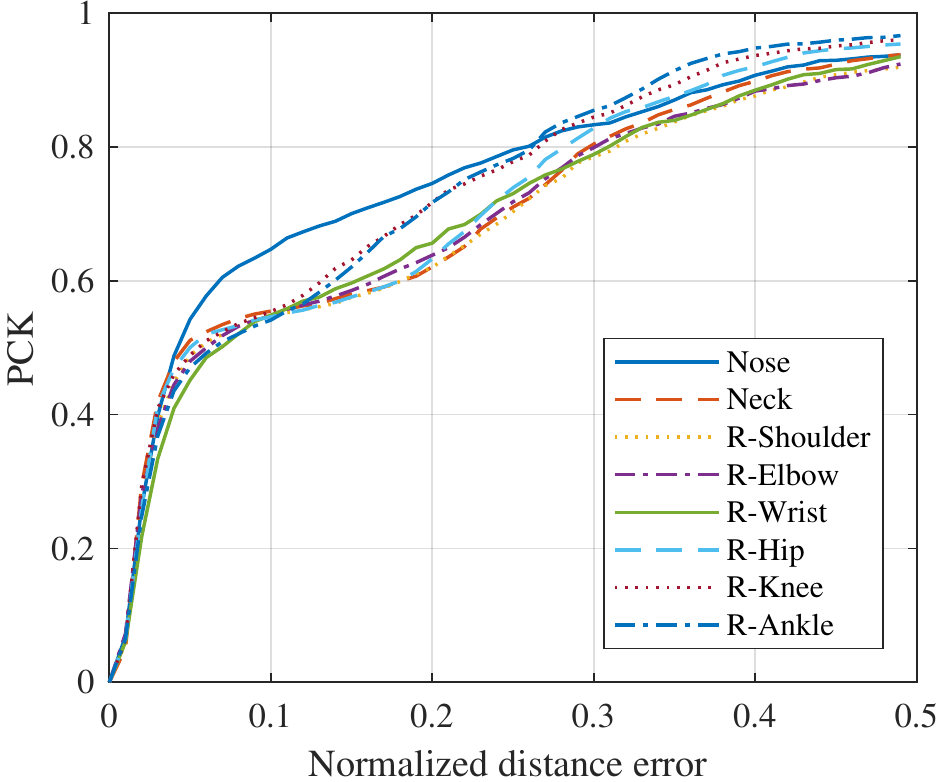}
		\caption{PCKs of CSI2Pose network. Left-side keypoints are not present for simplicity.}
		\label{fig:keypointpck}
	\end{minipage}
\end{figure*}

\subsection{Experimental Results}

\textbf{Forgery Detection Performance.} First, we present the detection performance of Secure-Pose. We report the receiver operating characteristic (ROC) curve of our system as well as its area under ROC curve (AUROC) to illustrate its capability of discriminating authentic and tampered video frames. As depicted in Fig.~\ref{fig:roccurve}, the ROC curve tightly follows the left-hand and top borders, implying that high TPRs and low FPRs are mostly obtained as the detection threshold varies in $ [0,1] $. Accordingly, the corresponding AUROC almost covers the whole ROC space and reaches to 0.983, which is close to 1, i.e., the ideal case. Furthermore, we demonstrate the detection performance under different forgery attacks in terms of TPR, FPR and accuracy. As shown in Fig.~\ref{fig:overallperformance}, Secure-Pose has a higher TPR and a lower FPR under inter-frame attacks. This is due to that inter-frame attacks require replacing all human objects in the authentic frame and thus incur a large difference between authentic and tampered frames, making our system easier to detect such forgeries. In contrast, intra-frame attacks only change a part of objects and therefore lead to a smaller difference. Despite that, our system still has a close performance in two attack scenarios. To sum up, Secure-Pose achieves an average detection accuracy of 95.1\% for each video frame. Specifically, it can successfully detect 94.9\% of tampered video frames and correctly recognize 95.3\% authentic frames. The above results demonstrate the effectiveness of Secure-Pose in detecting forgery attacks in video streams.

\begin{figure*}
	\hfill
	\begin{minipage}[t]{0.32\linewidth}
		\centering
		\includegraphics[width=1\linewidth]{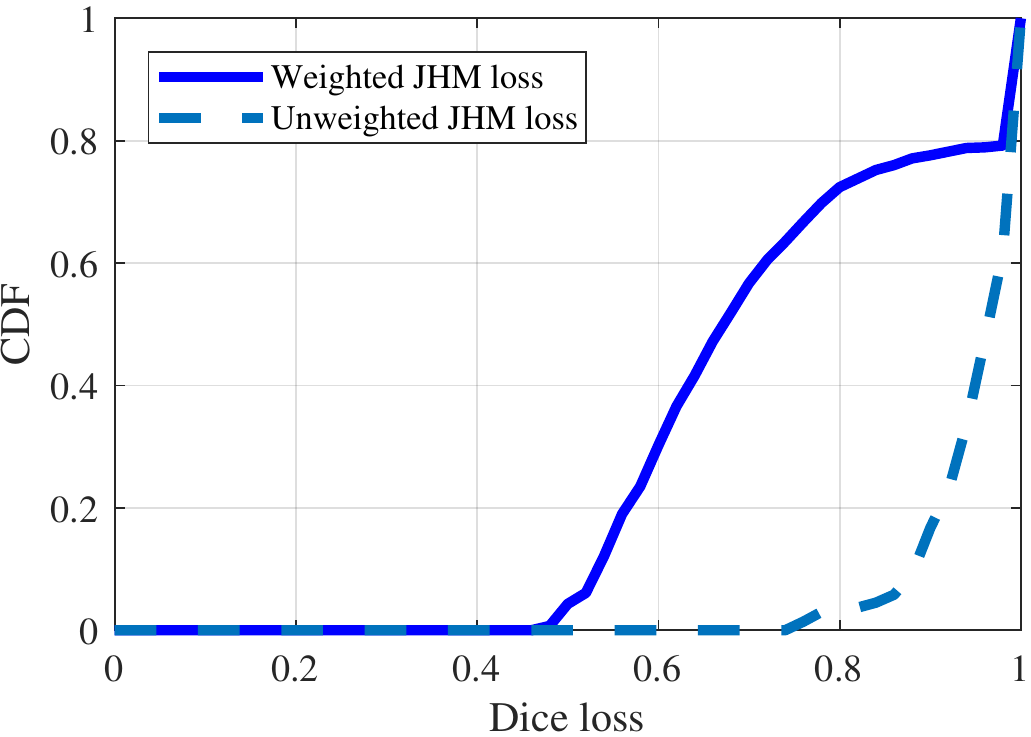}
		\caption{CDFs of Dice losses between CSI2Pose's and OpenPose's JHM outputs using weighted and unweighted losses.}
		\label{fig:cdfjhm}
	\end{minipage}
	\hfill
	\begin{minipage}[t]{0.32\linewidth}
		\centering
		\includegraphics[width=1\linewidth]{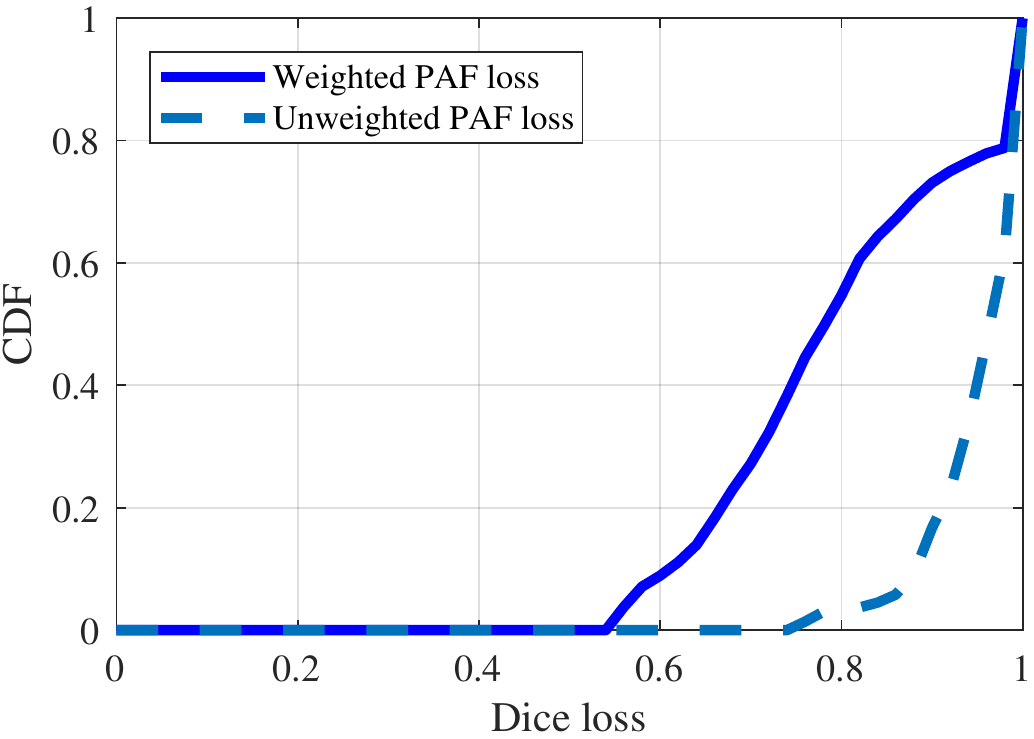}
		\caption{CDFs of Dice losses between CSI2Pose's and OpenPose's PAF outputs using weighted and unweighted losses.}
		\label{fig:cdfpaf}
	\end{minipage}
	\hfill
	\begin{minipage}[t]{0.32\linewidth}
		\centering
		\includegraphics[width=1\linewidth]{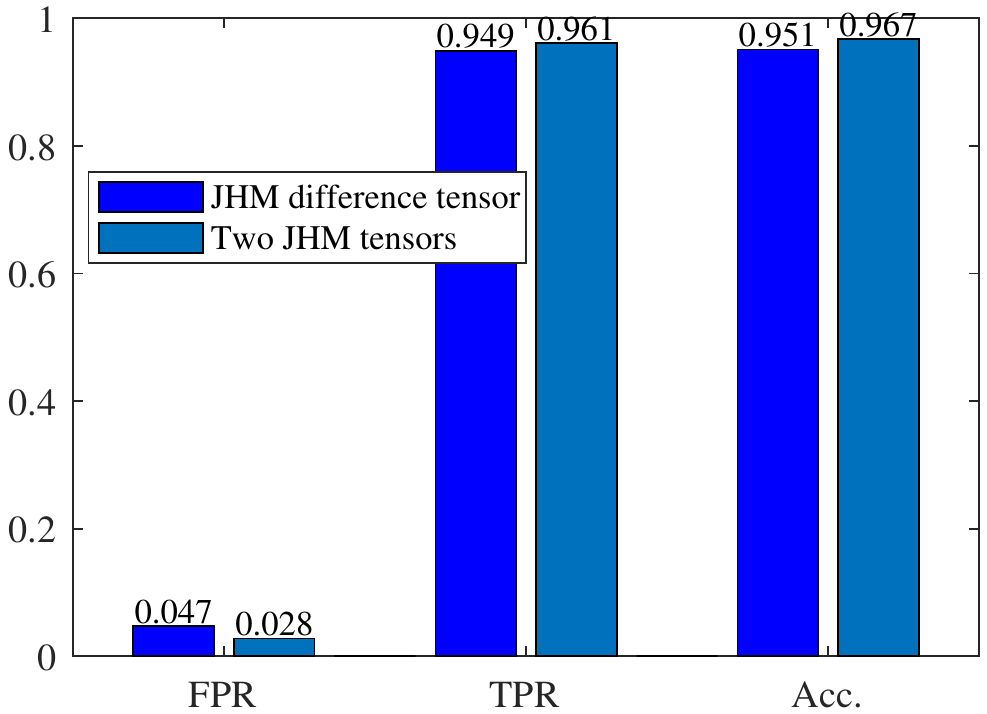}
		\caption{Performance of the detection network using different JHM features as input.}
		\label{fig:jhmdifferencetensor}
	\end{minipage}
\end{figure*}

\textbf{Forgery Localization Performance.} Then, we illustrate the forgery localization performance of our system. For this purpose, we show the mean PCK of pose estimation of abnormal objects under different forgery attacks. As depicted in Fig.~\ref{fig:forgerylocalizationperformance}, the localization of abnormal objects' keypoints yields high mean PCKs at low normalized distance errors. For instance, there are about 70\% estimated locations of abnormal keypoints that are within 20\% of diagonal length of person bounding boxes. Moreover, it also can be observed that our system always has a higher mean PCK under intra-frame attacks in terms of different normalized distance errors. The reason is that compared with inter-frame attacks, intra-frame attacks have a smaller modification on the difference JHM tensor $ \mathbf{D}^m $ as well as the combined PAF $ \mathbf{F}^m_{\mathbf{L}} $ in Eq.~\eqref{eq: combined paf}, which leads to a higher accuracy on keypoint localization and association for abnormal objects. Additionally, we showcase the outputs of forgery trace localization under inter-frame and intra-frame attacks, respectively, in Fig.~\ref{fig:instanceinterframe}. In the inter-frame attack, we replace the authentic frame that has three persons with a blank background. In this case, the absence of three persons is successfully detected by our system, and then their associated body keypoints are accurately inferred and presented on the tampered frame. In the intra-frame attack, one of two persons (the left one) is erased from the pristine frame. In this case, our system correctly localizes such forgery and plots the body pose of the erased person. The above results suggest the effectiveness of Secure-Pose in localizing different forgery attacks in each video frame.

\textbf{Performance of CSI2Pose Network.} Next, we show the performance of the proposed CSI2Pose network $ \mathcal{F_{W}} $. Specifically, $ \mathcal{F_{W}} $ is designed to infer human pose features from CSI signals with the supervision of OpenPose network $ \mathcal{F_{I}} $. To measure its ability of pose estimation, we directly leverage body association method~\cite{cao2017realtime} on CSI2Pose network's outputs $ \left( \mathbf{S}^m_{\mathbf{W}}, \mathbf{L}^m_{\mathbf{W}} \right)  $ and compare its association results with those of OpenPose network. As Fig.~\ref{fig:keypointpck} shows, the keypoint of nose has a higher PCK than the others at low normalized distance error. This may be caused by that the head part is generally with fewer occlusions by other people and clothing, which makes $ \mathcal{F_{I}} $ to generate more accurate nose labels for $ \mathcal{F_{W}} $ during cross-modal training and thus their performance difference of the nose part is smaller. Moreover, we also observe that as the normalized distance error approaches 0.5, the curves of all body keypoints get close with each other and tend to converge to a high value about 0.9. The above observations illustrate that the proposed CSI2Pose network is effective in semantic feature extraction.

\textbf{Effectiveness of the Weighted Loss Functions.} We further demonstrate the impact of the weighted MSE loss functions $ \mathcal{L}_{JHM} $ and $ \mathcal{L}_{PAF} $ proposed in Eq.~\eqref{eq:weighted jhm loss} and Eq.~\eqref{eq:weighted paf loss} for cross-modal training. To do this, we first train a CSI2Pose network using the standard MSE losses for JHMs and PAFs without weights as baseline. Then, we run both weighted and unweighted CSI2Pose networks on all testing samples and, respectively, compute Dice losses between their outputs and those of OpenPose network. Fig.~\ref{fig:cdfjhm} and Fig.~\ref{fig:cdfpaf} plot the cumulative distribution function (CDF) of Dice losses in terms of JHM and PAF, respectively. In Fig.~\ref{fig:cdfjhm}, the weighted curve is always higher than the unweighted one. Specifically, the unweighted curve is very steady and goes straight up when the Dice loss exceeds 0.8, however, the weighted curve rises up beyond 0.5. The same trends can also be found in Fig.~\ref{fig:cdfpaf} for PAFs. The above observations indicate that the weighted losses $ \mathcal{L}_{JHM} $ and $ \mathcal{L}_{PAF} $ are effective, and they can force CSI2Pose network to pay more attention on foreground objects and thus learn more accurate JHM and PAF features from WiFi CSI features during cross-modal training.

\textbf{Effectiveness of the JHM Difference.} Finally, we present the effectiveness of the proposed JHM difference tensor $ \mathbf{D}^m  $, which is the difference between estimated JHM tensors $ \mathbf{S}^m_{\mathbf{I}} $ and $ \mathbf{S}^m_{\mathbf{W}} $. To do this, we modify the detection network $ \mathcal{F_D} (\cdot) $ to take together $ \mathbf{S}^m_{\mathbf{I}} $ and $ \mathbf{S}^m_{\mathbf{W}} $ as input and train it on Dataset A. Fig.~\ref{fig:jhmdifferencetensor} illustrates TPR, FPR and accuracy of two versions of $ \mathcal{F_D} (\cdot) $ using different JHM feature tensors. We can observe that two complete JHM tensors enable a TPR increase of 1.2\% and bring a FPR decrease of 1.9\% in comparison with those of the JHM difference tensor. This observation is reasonable because $ \mathbf{S}^m_{\mathbf{I}} $ and $ \mathbf{S}^m_{\mathbf{W}} $ naturally have more formation about forgery attacks than $ \mathbf{D}^m  $, which can be learned by our detection network. Despite that, they have nearly identical performance. The above observations suggest that the proposed JHM difference tensor $ \mathbf{D}^m  $ is an effective feature representation for forgery detection. In particular, it can preserve suffice information about forgery traces that is contained in JHM tensors $ \mathbf{S}^m_{\mathbf{I}} $ and $ \mathbf{S}^m_{\mathbf{W}} $ while having a smaller size.

\section{Related Work}\label{sec:related work}
\textbf{Video Forgery Detection.} With the rapid advances in video editing techniques, it has become much easier to tamper surveillance videos. Traditional watermark-based approaches require dedicated modules for video integrity preservation, while not all commercial cameras have such a watermarking module. Passive video forensics approaches leverage video statistic characteristics to discover forgery traces~\cite{fayyaz2020improved,yang2016using,chen2015automatic,ulutas2017frame,wang2007exposing}. However, such approaches are computationally-expensive and cannot be applied for real-time forgery detection on live video feeds. The recent work~\cite{lakshmanan2019surfi} compares event-level timing information from WiFi and camera signals to detect camera looping attacks. However, this work cannot detect and localize forgery attacks in each video frame. In this work, we propose a novel cross-modal system that can detect and localize forgery attacks in each frame of live surveillance videos.

\textbf{Human Perception Using RF Signals.} Recent years have witnessed much progress in performing human perception based on RF signals. These studies~\cite{wang2018securing,luo2018authenticating,yong2020authenticating} exploit unique RF characteristics to detect on-body propagation. Moreover, the authors in~\cite{adib2015capturing} use a dedicated FMCW radio to capture coarse human skeletons from RF signals that bounce off the body when a subject walks towards transceivers. After that, they proceed to exploit FMCW radios to estimate 2D~\cite{zhao2018through} and 3D \cite{zhao2018rf} human poses through walls and occlusions. Compared with dedicated RF radios, low-cost WiFi radios are much promising in enabling pervasive human perception in various indoor environments. Relying on commercial WiFi transceivers, WiPose~\cite{jiang2020towards} is proposed to reconstruct 3D human skeletons in the single-person scenario. Moreover, Person-in-WiFi~\cite{wang2019person} is designed to learn body segmentation mask and joint coordinates from WiFi signals. In this work, we focus on recover human pose features from both video and WiFi signals and perform video forgery detection and localization in a cross-modal manner.

\section{Conclusion}\label{sec:conclusion}
This paper presents Secure-Pose, a novel cross-modal system that effectively detects and localizes forgery traces in live surveillance videos using ambient WiFi signals. We observe that the coexisting camera and WiFi signals contain common human semantic information and the presence of video forgery attacks will decouple such cross-modal information correspondence. Our system effectively extracts human pose features from synchronized camera and WiFi signals and efficiently discovers forgery traces under different attacks. We implement our system using one commercial camera and two Intel 5300 NICs and evaluate it in real-world environments. The evaluation results demonstrate that Secure-Pose achieves a high detection accuracy of 95.1\% and accurately localizes body keypoints of tampered objects under both inter-frame and intra-frame attacks.

\section*{Acknowledgement}
This work was supported in part by the National Key R\&D Program of China under Grant 2020YFB1806606, National Science Foundation of China with Grant 62071194, 61729101, 91738202, Young Elite Scientists Sponsorship Program by CAST under Grant 2018QNRC001.

\balance

\bibliographystyle{IEEEtran}
\bibliography{IEEEabrv,./Forgerydetection}

\end{document}